\documentclass[12pt,preprint]{aastex}

\shorttitle{CXO/XMM Imaging of GeV J1417$-$6100}
\shortauthors{}

\begin{document}

\title{Two Pulsar Wind Nebulae: Chandra/XMM-Newton Imaging of GeV J1417$-$6100}

\author{C.-Y. Ng$^1$, Mallory S.E. Roberts$^2$, \& Roger W. Romani$^1$}
\affil{$^1$Department of Physics, Stanford University, Stanford, CA 94305}
\affil{$^2$Department of Physics, McGill University, 3600 University,
Montr\'eal, QC H3A 2T8 Canada}
\email{ncy@astro.stanford.edu, roberts@physics.mcgill.ca, rwr@astro.stanford.edu}

\begin{abstract}

We report on {\it Chandra} ACIS and {\it XMM-Newton} MOS/PN imaging observations of two
pulsar wind nebulae (K3/PSR J1420$-$6048 and G313.3+0.1=`the rabbit') 
associated with the Galactic unidentified $\gamma$-ray source GeV J1417$-$6100.
With the excellent {ACIS} imaging the very energetic pulsar PSR J1420$-$6048 
is separated from its surrounding nebula. This nebula has surprisingly little
compact structure, although a faint arc is seen near the
pulsar. Similarly, two point sources are resolved in the 
rabbit nebula. The large {\it XMM-Newton} collecting area provides useful spectral
constraints on the rabbit and the associated point sources. Based on 
spectra and X-ray morphology, we identify one point source as a plausible 
pulsar counterpart. Large backgrounds and low source counts limited pulse
search sensitivities, but we report pulse upper limits and a candidate
108ms period for the rabbit pulsar based on the {\it XMM-Newton} data and an
{ACIS} CC observation.
Comparison of the X-ray images with high resolution ATCA radio
maps shows that the non-thermal X-ray emission corresponds well with
the radio structure. 
\end{abstract}


\section{Introduction}

	The {\it EGRET} instrument on the {\it Compton Gamma Ray Observatory}
detected a number of sources along the Galactic plane which remain 
unidentified. The sources which are strongly detected at $E> $GeV
energies (Lamb \& Macomb 1997) are the best localized and X-ray
studies (Roberts, Romani \& Kawai 2001) provide a list of possible
counterparts. GeV J1417$-$6100 is one of the brightest unidentified GeV
sources and is particularly
interesting since two apparent pulsar wind nebulae lie within its
95\%-CL uncertainty region (Roberts, Romani, Johnston \& Green 2001;
Roberts, Romani \& Johnston 2001, RRJ). 

	The radio/X-ray morphology of this region is complex. A thermal
shell is flanked by two non-thermal radio wings. The northern wing is
well centered in the GeV 68\% uncertainty contour and contains non-thermal
X-rays. A young $\tau_c=1.3 \times 10^4$y pulsar, PSR J1420$-$6048, was 
found in this region during the Parkes multibeam survey (D'Amico et al 2001).
With a spindown luminosity ${\dot E} = 1 \times 10^{37}$erg/s, this is
the tenth most energetic radio pulsar known.  RRJ studied this pulsar
with an {\it ASCA} pointing, {\it ATCA} observations at 20cm and 13cm,
and Parkes 20cm radio pulse and polarization observations. The {\it ASCA}
data provided a marginal detection of X-ray pulsations, but suggested that
the majority of the flux came from extended diffuse emission. However,
the modest ($\sim 2^\prime$ FWHM)
spatial resolution of the {\it ASCA} data made it difficult to
disentangle extended nebulae from embedded point sources.

To firmly identify PSR J1420$-$6048 as a gamma-ray pulsar would
require the detection of $\gamma$-ray pulsations. Radio
timing, however, shows that the pulsar glitches and so it is perhaps
not surprising that extrapolation of the pulsar ephemeris to the {\it EGRET}
epoch has not provided a pulse detection.  It is important to note that the
gamma-ray source shows evidence of variability \citep{noet03, cb99},
suggesting the source is not 100\% pulsed and that 
there is a significant contribution from a pulsar wind nebula (PWN) shock.
In this connection it is also interesting that 15$^\prime$ to the SW
lies another, brighter non-thermal X-ray/radio nebula, referred to
in RRJG as the `rabbit'. Indeed this source is also within the {\it EGRET}
source 95\% positional uncertainty region, which has an extension toward
the southwest. This suggests that more than one source may contribute to the
GeV emission.
The rabbit nebula has the radio/X-ray characteristics, including high
radio polarization, of a PWN, although again limited {\it ASCA} resolution
precluded study of its X-ray structure. Here we report
on {ACIS} and {\it XMM-Newton} observations that allow a study of these nebulae
and their central compact objects.

\section{Observations and Data Analysis}

\subsection{X-ray Imaging and Source Variability}

PSR J1420$-$6048 was observed on 16 September 2002 (MJD 52533, ObsID 2792) using the
ACIS-S array for 10~ks in VF TE imaging mode. During this observation, the linear array
of 4 ACIS-S chips was rolled to provide off-axis coverage of the rabbit, while the I0
and I1 chips provided coverage of two point sources seen in the $ASCA$ observations.
Immediately after this exposure, we obtained 31~ks of CC observation with the aimpoint
shifted to mid-way between PSR J1420$-$6048 and the rabbit nebula
(ObsID 2793).  The program was completed on 22 September 2002 (MJD 52539, ObsID 2794)
with a 10~ks VF TE exposure with the ACIS-S3 aim-point now directed to near the rabbit
nebula, with the ACIS-S array at a similar roll angle providing off-axis coverage of
PSR J1420$-$6048. The data were subject to standard calibrations with CIAO version 3.1.
Examination of the background count-rates during these exposures showed no strong
flares, so all on-source data were included in the analysis. The rabbit was also
observed with {\it XMM-Newton} for $\sim25$~ks on 10 March 2003 (MJD 52708). For this
observation, the PN camera was in small window mode to allow a
search for pulsations from point sources seen in the rabbit, and the medium filter
was used with the MOS cameras.  The SAS Version 10040728 was used for the data
calibration, with times of high background flaring removed. Figure 1 shows the
$Chandra$ and {\it XMM-Newton} imaging data, exposure weighted and smoothed, with contours are 
from the 20cm ATCA radio image of RRJ01; the shell and extensions were there referred to as
the `Kookaburra'. The location of the brighter sources discussed
in this paper are indicated.


\begin{figure}[htb]
\epsscale{1} \plottwo{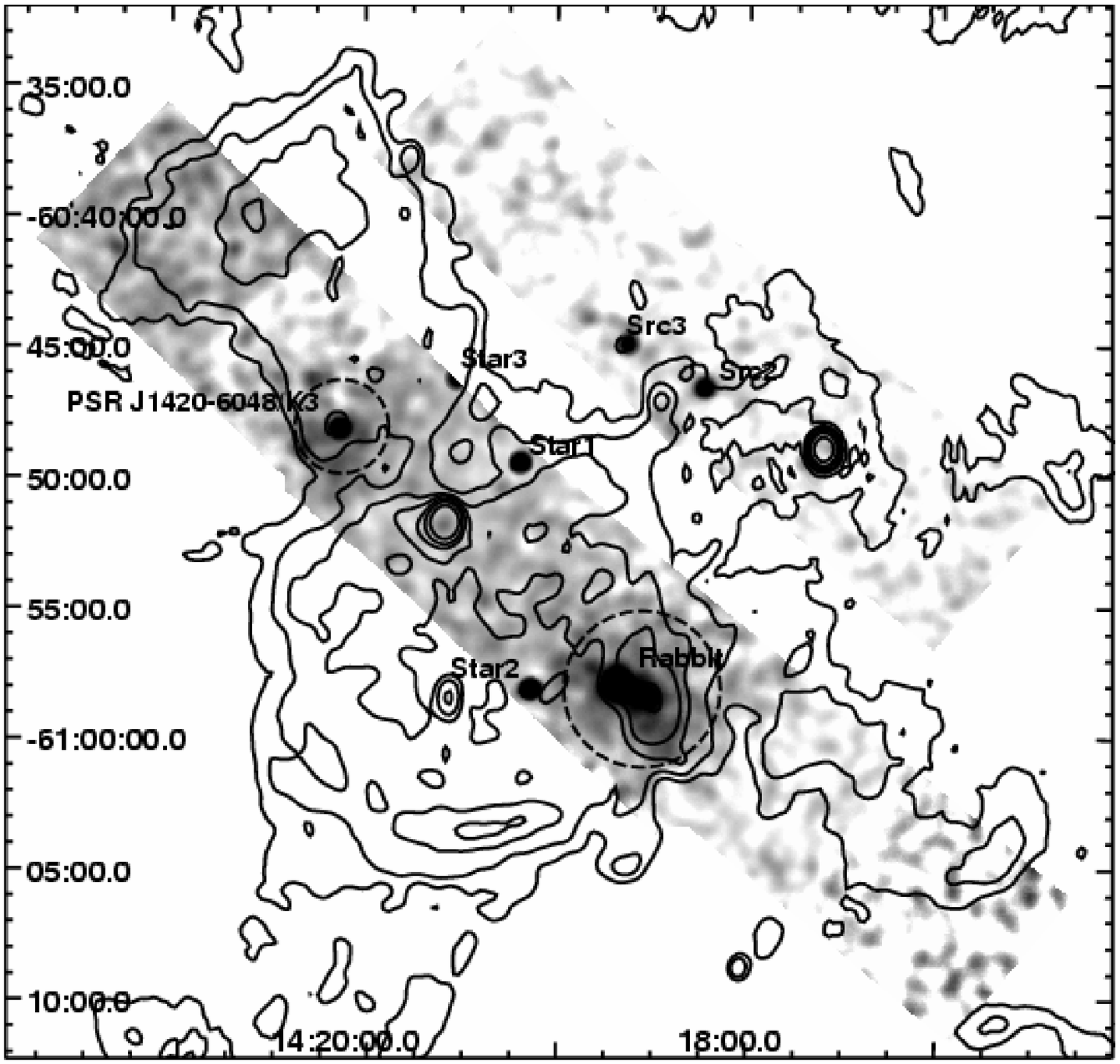}{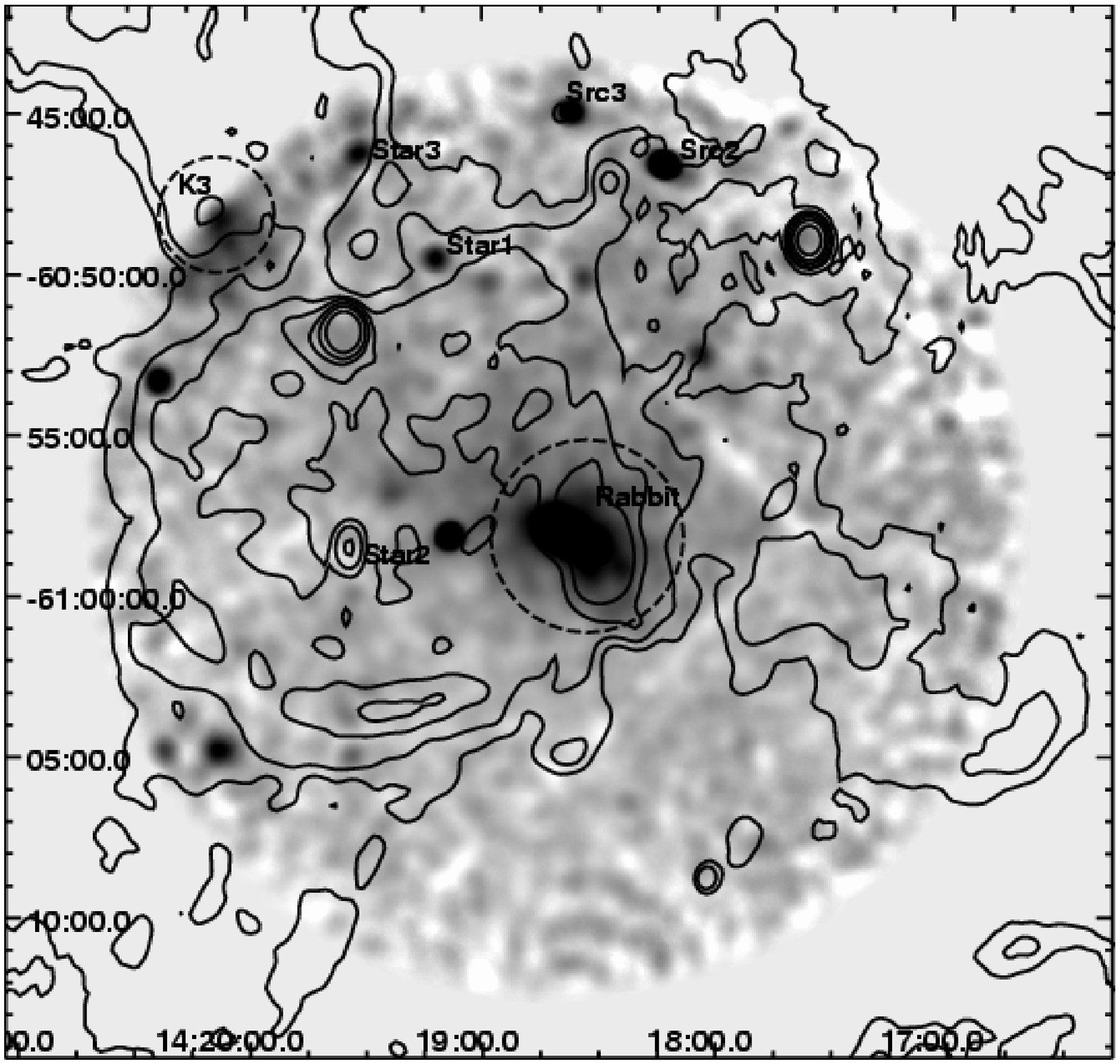}
\caption{Smoothed X-ray images of the `Kookaburra' complex. Left: our combined ACIS TE exposures 
(1-7keV, $15^{\prime\prime}$ Gaussian smoothing), Right: our MOS1/MOS2 image (1-7keV,
$15^{\prime\prime}$ Gaussian smoothing). Both images show overlays from the
20 cm ATCA radio contours (5,10,20,30,50,80,160 mJy/bm).
The brightest X-ray sources common to both observations are labeled.
We also plot dashed line circles marking the K3 PWN ($2^\prime$ radius) and the 
`Rabbit' PWN ($3^\prime$ radius).}
\label{xray-radio}
\end{figure}


Three of the brightest X-ray sources are soft (few counts above 2keV)
and clearly identified 
with bright field stars (Star1-3). Two bright hard X-ray sources (Src2, Src3), 
suggested by RRJG to be AGN viewed through the plane, show no optical/IR 
counterparts.  Only Src3 is clearly detected in the radio as a flat
spectrum, variable point source with typical flux of a few mJy.



\begin{figure}[htb]
\epsscale{0.5} \plotone{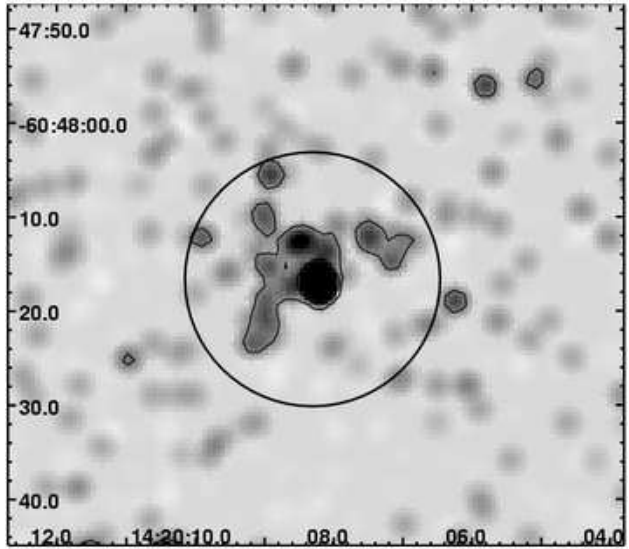}
\caption{The inner K3 nebula, which is embedded in faint diffuse
emission (Figure 1). An arc of X-ray photons partly surrounds the point 
source, PSR J1420$-$6048. The 13$\farcs 5$ circle
show the extraction region for spectra of the inner nebula.}
\label{K3}
\end{figure}


	The high angular resolution of the {\it Chandra} ACIS allows us to
investigate point sources in the PWNe. Figure 2 shows a portion of ObsID 2972
in which PSR J1420$-$6048 is clearly detected with 26 counts (see Table 1). 
The nebular emission closest to the pulsar seems to form a faint arc containing
$\sim 55$ counts, running East through North.
The closest approach, at PA$\approx45^\circ$, is 
$\sim 3^{\prime\prime}$. Within the rabbit nebula, there are two unresolved point 
sources (R1, R2). A third cluster of photons to the north of these two sources 
is unresolved in the $XMM-Newton$ images but is clearly extended in the 
on-axis ACIS image with a half-power width of $5^{\prime\prime}$, 10$\times$
that of the local PSF.  Table~1 gives
the ACIS positions and 1-8 keV count-rates (exposure and aperture corrected)
for our bright sources in the two TE exposures. We also tested the sources for
variability during each exposure using the K-S statistic (Press et al. 1987)
by comparing the individual source photon arrival times to those from the background.
Only Star1 shows clear variability during an observation, with a strong flare
during the second half of ObsID 2792. Source R1 shows a $1.5\sigma$ change in
count-rate between TE exposures, while Src2 shows a $\sim 3\sigma$ change in
count-rate. R2 shows weak 90\% evidence for variability during each exposure,
although the low count-rate is consistent between observations. All
other sources are consistent with steady emission. In the $XMM-Newton$ data
R1 and R2 are consistent with steady flux.

Extra care needed to be taken with the imaging of the nebulae, since the extended
emission has such low surface brightness. Images were extracted in 12 energy bands
between 0.5 and 10 keV from the TE data and the stowed data set
available from the CXC. The images from the stowed data, which reflect the
particle background spatial distribution and spectrum, were scaled by the
relative number of counts at energies where the particle background dominates
and then subtracted from the source images. Each was then individually exposure
corrected using a single energy exposure map and the resulting images summed.
Examination of the images showed little evidence for the nebulae outside the
1-7 keV range, and so that range was used for the spatial analysis. A similar
procedure was followed with the {\it XMM-Newton} data.  These images were then 
smoothed with Gaussian kernels of various widths in order to bring out 
structure on various spatial scales.

At the broadest scales, the majority of the flux from both of the PWNe is in
large low-surface brightness nebulae with indistinct morphologies (Figure 1). 
The $Chandra$ image of the diffuse emission around PSR J1420$-$6048
shows the pulsar to be near the north edge of a region of emission $\sim 3^{\prime}$
across. Except for the small arc noted above, there are no other distinct structures
at other scales, although $ASCA$ data suggest fainter emission on even larger scales. 
The rabbit (Fig. 3) has a clear brighter region at
DEC=-60:58 extending from the cluster of photons just north of the two point
sources toward the southwest, correlated with the rabbit radio nebula. In the
radio, this region shows the strongest polarization and spectral tomographic
maps show a distinctly flatter spectrum than for the ring forming the `body' of the
kookaburra. The X-ray data shows that the body of the rabbit is clearly distinct,
and likely unrelated to, the thermal shell.  However, there is also much 
fainter emission nearly filling the central chip of the $XMM-Newton$ MOS image
and extending $\sim 4^\prime$ from the central point sources.

In Figure 3 we show the brighter region of the X-ray rabbit with
moderate smoothing. The overlay contours are from a high resolution 20~cm $ATCA$
image of the rabbit. The brighter point source, labeled R1, lies at the edge of
the rabbit, while the fainter source (R2) appears embedded in the
diffuse emission. It is presently unclear which, if either,
of these sources is the nebula's parent pulsar, although timing and spectroscopic
evidence below suggests an association with R2.
The general impression is that there is a sub-luminous zone 
around R2 of radius $\sim 25^{\prime \prime}$, bordered by an arc of
diffuse emission. The diffusion emission continues out to near $5^\prime$,
with low surface brightness. The brightest core of the rabbit seems
to stretch $\sim 3^\prime$ at a position angle $\sim 230^{\circ}$, 
measured N through E. This structure correlates well with the flat spectrum 
radio emission.


\begin{figure}[htb]
\epsscale{1} \plotone{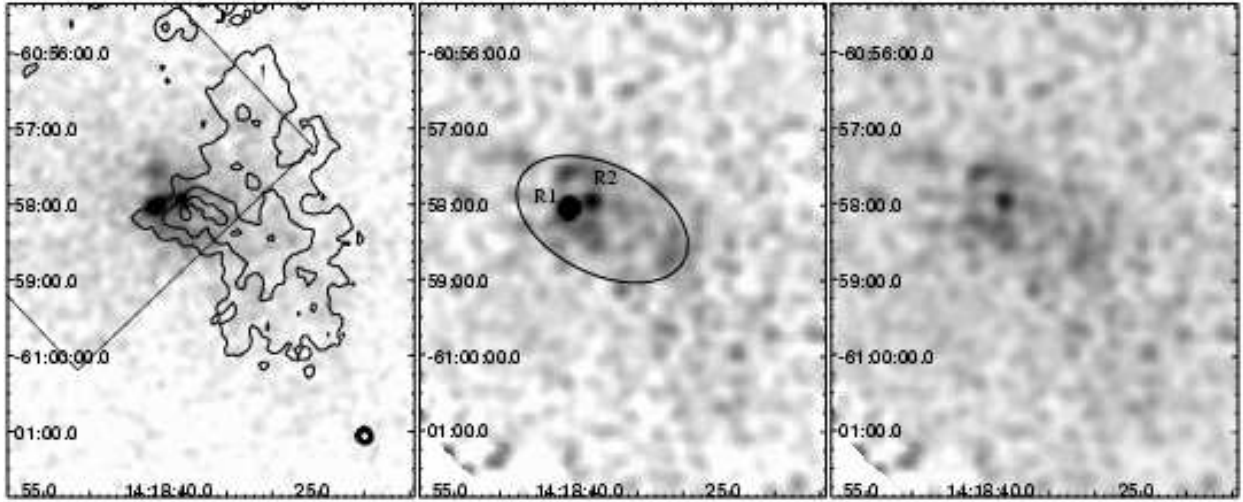}
\caption{The rabbit nebula. Left: combined, exposure corrected 1-7KeV MOS1,2 
and PN camera images, smoothed with a $4\farcs9$ Gaussian kernel. 
The square shows the region also covered by the PN camera
in small window mode. The contours are drawn from the high resolution
(6km baseline) 20cm ATCA map of RRJG. Middle: 1-7keV {\it ACIS} TE imaging 
of the rabbit, exposure corrected
and smoothed with a $4\farcs9$ Gaussian kernel. The ellipse shows the spectral
extraction region for the nebula in the {\it ACIS} data. 
Right: 1-7keV smoothed image after removal of the point source 
R1, showing diffuse emission symmetric about the source R2.}
\label{threepan}
\end{figure}


\begin{table}[h]
\tablecaption{{\it Chandra} Bright Source Positions, Counts and Variability.}
\begin{tabular}{l|cc|ccc|ccc}
& & & \multicolumn{3}{c|}{obs-id 2792} & \multicolumn{3}{c}{obs-id 2794} \\
& & & Aper. & & Corrected & Aper. & & Corrected \\
Object & RA & Dec & cnts\footnotemark[1] & Prob.\footnotemark[2] & flux\footnotemark[3] 
& cnts\footnotemark[1] & Prob.\footnotemark[2]  & flux\footnotemark[3] \\
\hline \hline
PSR J1420 & 14:20:08.20(8) & -60:48:17.2(5) & 26 & 0.56 & $4.6\pm 1.0
$ & 21 & 0.05 & $5.3\pm 1.8$ \\
R1 & 14:18:42.70(8) & -60:58:03.1(4) & 22 & 0.76 & $4.4\pm 2.2$ & 
54 & 0.15 & $10\pm 1$ \\
R2 & 14:18:39.90(6) & -60:57:56.5(3) & 26 & 0.90 & $4.8\pm 2.6$ & 
20 & 0.90 & $3.5\pm 0.9$ \\
Src 2 & 14:18:13.42(7) & -60:46:41.5(6) & 55 & 0.53 & $15\pm 2$
 & 36 & 0.14 & $8.7\pm 1.6$ \\
Src 3 & 14:18:37.83(7) & -60:45:01.1(1) & 43 & 0.89 & $20\pm 2$
 & - & - & - \\
Star 1 & 14:19:11.75(6) & -60:49:33.8(7) & 187 & 0.70 & $42\pm 3
$ & 337 & 0.9999 & $87\pm 5$ \\
Star 2 & 14:19:08.28(9) & -60:58:13.9(6) & 121 & 0.19 & $24\pm 3
$ & 162 & 0.09 & $27\pm 2$ \\
Star 3 & 14:19:31.65(9) & -60:46:21.3(3) & 143 & 0.11 & $24\pm 2
$ & 70 & 0.12 & $19\pm 3$ \\
\end{tabular}
\label{var}
$^1$0.5-8 keV counts in extraction apertures matched to the model PSF EE80 ellipse.

\qquad $^2$Probability of significant variability (KS statistic).
\qquad $^3$in units of cts/Ms/cm$^2$, corrected for background and aperture losses.
\end{table}

\subsection{Timing Analysis}

We have attempted to detect the pulsations of PSR J1420$-$6048 in our 
continuous clocking observation (obs-id 2793). Inspection of the 1-D
collapse of the chips shows a clear detection of PSR J1420$-$6048, but it
competes with a large background, both from the surrounding nebulae and 
from diffuse emission across the chip.

In the continuous clocking observation, the arrival times of the photons 
are calculated from the readout time (2.85ms resolution) and then barycenter 
corrected using CIAO.  We extract 261 counts in the 1-7 keV band from a 
2.5$^{\prime\prime}$ radius aperture centered on the pulsar.
However, an estimate of the flanking background implies that only $\sim$38\% 
of the counts are from the point source. We folded the CC data on
an updated radio ephemeris covering the {\it Chandra} observations
(courtesy George Hobbs) which predicts $P=0.0681865728$s,
${\dot P} = 8.2807487 \times 10^{-14}$ at epoch MJD 52534.0.
Fitting a sinusoidal pulse plus constant to a 10 bin light curve
gives a pulse fraction of $PF=0.55\pm 0.43$, providing no significant
constraint on the X-ray pulsations. Similarly a check for
periodicity using the $H$-test (deJager, Swanepoel \& Raubenheimer 1989) 
obtains a value $H=0.89$; the chance probability of larger $H$ is 70\%. 
The large background and the lower than expected source count-rate thus preclude
a sensitive pulse search in these CC data. PSR J1420$-$6048 was not covered by
the {\it XMM-Newton} exposure.

	We also searched the point sources in the rabbit for pulsations,
starting with the (relatively) high count-rate EPIC-PN data. We determined
that R2 dominated the diffuse background out to a $6^{\prime \prime}$ radius
at energies $\sim 1.3-6.4$\,keV. These cuts provided 132 counts from the PN
detector. An H-test scan at 1/5 Nyquist frequency resolution from 0.05-0.5\,s 
(4762279 trials) found three periods with chance probability $\le 0.5$, after
correcting for the number of trials.  A similar scan of counts extracted from 
R1 provided no significant periods.  For the expected PSF significant
counts should extend to $10^{\prime \prime}$. Of the three periods found for R2 
only one ($P=0.108$\,s) with chance probability 0.25 retained high single trial
significance for all three apertures. A fold of these photons resulted in 
a light curve with a single narrow peak. We then searched photons form
the R2 ACIS-CC
data. This source was $\sim 8.5^{\prime}$ off-axis, requiring a  $\sim 
5^{\prime \prime}$ half-power extraction radius. This aperture provided 427
source and background counts, using the same energy range as for {\it XMM-Newton}.
Unfortunately, at the observation roll angle, the projected separation from R1 
was only $8^{\prime \prime}$, and so this source contributed to the already
substantial background. We searched the $\dot P$ range corresponding to
${\dot E_{SD}} = 10^{35}-10^{38}$\,erg/s (1298 at 1/5 Nyquist sampling); the
most significant value was $\dot P=1.0688\times 10^{-12}$ with a trial-corrected
significance of 0.82. Finally, we searched the high-bit rate GIS {\it ASCA}  data of RRJ
(February 13-15 1999, MJD=51222-51224),
which provided 1820 counts within the half 
power diameter, over a small $\dot P$ range about the {\it Chandra/XMM-Newton} value.
There is a peak at $\dot P=1.0547e-12$, within 1.5\% of the {\it Chandra/XMM-Newton} value
and well within the range allowed by timing noise for such a young pulsar.
For the 607 trials within 1.5\%, the peak has a significance of 0.66. The fold
of the photons at the epoch period produces a light curve similar to that
from the {\it XMM-Newton} data.
The combination of these measurements gives a joint probability of chance
detection of 0.02, so the detection is intriguing, but not highly significant.

\subsection{Spectral Analysis}

	We have analyzed the spectra of our brightest point sources and
our two extended nebulae.  All {\it Chandra} spectral analysis is done with CIAO 3.1 
and CALDB 2.26, which take care of the ACIS QE degradation automatically, 
hence no additional correction for the ACIS contamination is necessary.  
For the {\it XMM-Newton} data, spectra were fitted to the calibrated data sets using
the XSPEC package. Scaled closed-door spectra were subtracted from the
extraction regions to remove particle background. Analysis of the low 
signal-to-noise, extended flux 
in the rabbit depends critically on the background subtraction. In the MOS data
only a small background region was available on the central chip, so it was 
supplemented by photons from from another chip. {\it XMM-Newton} nebular results are
from the MOS data only as there was essentially no off-source region in the PN data.
With the low count-rates obtained for all sources
pileup is negligible. Given the limited statistics, only
simple absorbed power-law fits were attempted for all non-thermal sources;
Mekal thermal plasma fits were applied to the bright stars. 
The results are listed in Table 2. For consistency, all fits are to the 0.5-8\,keV
range. For comparison with scaling laws in the literature we also quote
unabsorbed 2-10\,keV fluxes. Spectral parameter errors are projected 
multi-dimensional $1\sigma$ values. As is often the case with low statistics
X-ray spectra, projected (multi-dimensional) errors on the fluxes are
very large due to spectral parameter uncertainties. Thus, we follow other
authors in quoting flux errors as $1\sigma$ single parameter values.

	A power-law fit to the K3 nebula as a whole finds a rather large
$N_H \approx  5.4 \times 10^{22} {\rm cm^{-2}}$
absorption column, albeit with large errors. The inner area of the nebula
(Figure 2) has a distinctly flatter spectrum, again with quite large errors.
Note that the pulsar dispersion measure DM=360${\rm cm^{-3}pc}$ 
(D'Amico et al 2001) implies an H column $N_H \approx 3.7 \times 10^{22} 
(n_H/30n_e) {\rm cm^{-2}}$. 
Although this is larger than the nominal total
H column in this direction ($\sim 2 \times 10^{22}{\rm cm^{-2}}$, FTOOLS
nh tool), it is reasonable for the total Galactic absorption to be this large.
Src3, whose identification with a flat spectrum radio point
source supports an extragalactic interpretation, is found in the {\it XMM-Newton}
fits to have a large absorbing column $N_H > 9 \times 10^{22} {\rm cm^{-2}}$,
although part of this absorption may well be internal to the source.
Even if we fix the X-ray spectral index to match a typical AGN spectrum of 
$\Gamma=1.7$, we still obtain $N_H > 5.3 \times 10^{22} {\rm cm^{-2}}$ for the 
{\it XMM-Newton} data and $N_H \approx 3.6 \times 10^{22} {\rm cm^{-2}}$ for the 
{\it Chandra} observations of this source. So a large $N_H$ is plausible in this
direction. Of course if we fix at lower $N_H$ in the fit, the inferred spectral
index of the K3 region hardens from its unconstrained-fit value $\Gamma=2.3\pm 0.9$ 
to $\Gamma = 1.5 \pm 0.3$ ($3.7\times 10^{22}$) and 
$\Gamma = 0.7 \pm 0.3$ ($2.0\times 10^{22}$). The fluxes do not change significantly.

For the on-axis {\it Chandra} observation of PSR J1420$-$6048 we find only 26
counts from the 0.5-8 keV energy band within an aperture of radius 
1$^{\prime\prime}$.  Fixing $N_H = 5.4\times 10^{22}$cm$^{-2}$ at the K3 value,
we find $\Gamma=1.0$ with very large uncertainty. Again the nominal index decreases
if we assume lower $N_H$ (0.7 at $3.7\times 10^{22}$; 0.4 at $2.0\times 10^{22}$)
but these are not statistically well constrained.
The absorbed flux is only 7$\times 10^{-14}$ ergs cm$^{-2}$s$^{-1}$.
Unfortunately the off-axis observation suffered from a large PSF
and background; thus a simultaneous fit does not improve the error bars.

	The rabbit nebula is somewhat brighter and is well fitted by an
absorbed power-law. In the {\it XMM-Newton} data we find $N_H = 1.4 \times 10^{22}
{\rm cm^{-2}}$, and $\Gamma = 1.5$ for a large extraction region that includes much
of the diffuse outer nebula. The brighter central region shows a larger 
$N_H=2.3\pm0.2 \times 10^{22} {\rm cm^{-2}}$ and $\Gamma=1.8\pm0.14$.
A free fit to the bright diffuse emission of the central nebula in the 
combined ACIS TE exposures gives also somewhat larger $N_H$ and $\Gamma$, 
but if $N_H$ is held at the {\it XMM-Newton} value for the inner nebula the power-law index 
$\Gamma=1.7$ is close to the {\it XMM-Newton} fit value. The two embedded point sources 
are reasonably hard, as well. In the {\it XMM-Newton} data, we find a column 
$N_H = 2.0 \times 10^{22} {\rm cm^{-2}}$ for R1, consistent with the rabbit nebula.
R2 shows $N_H = 2.5 \times 10^{22} {\rm cm^{-2}}$ nominally higher than,
but still consistent with, the nebular value.  The {\it Chandra} observations give too 
few counts for sensitive fits, but the nominal values are not inconsistent 
with the {\it XMM-Newton} parameters.

	Of the other bright sources, we find that Src2 provides a rather poor
fit to a power-law. In the two {\it Chandra} TE observations, we find that the
spectrum varies substantially. However, if a simple power-law fit is forced,
both the {\it Chandra} and {\it XMM-Newton} data give a modest $N_H \sim 10^{22}
{\rm cm^{-2}}$. This raises the possibility that the source is Galactic,
or more likely that it has a composite spectrum. The stars are all consistent
with low absorptions and relatively soft, thermal spectra. Star1, which
shows clear evidence for a flare during ObsID 2794, is not unexpectedly
fitted with a higher effective temperature and flux during this observation.

\begin{table}[!t]
\caption{Spectral Fits}
\begin{tabular}{l|ccc|ccc}
& \multicolumn{3}{c|}{\emph{Chandra}} & \multicolumn{3}{c}{\emph{XMM-Newton}} \\
object & N$_H$\footnotemark[1] & $\Gamma$/kT\footnotemark[2] & flux\footnotemark[3] & N$_H$\footnotemark[1
] & $\Gamma$/kT\footnotemark[2] & flux\footnotemark[3] \\
\hline
K3 & $5.4^{+2.2}_{-1.7}$ & $2.3^{+0.9}_{-0.8}$ & $6.7\pm0.7$(13$\pm 1.4$)\\
K3 (inner) & $5.4^*$ & $0.5^{+1.3}_{-1.1}$ & $1.3\pm0.3$(2.4$\pm 0.6$) \\
PSR J1420 & $5.4^*$  & $1.0^{+4.2}_{-4.8}$ & $0.7\pm0.3$(1.3$\pm 0.6$) \\

\hline
Rabbit (full)&  &  & & $1.4\pm0.2$ & $1.5\pm0.14$ & $63\pm 2$($73\pm2 $) \\
Rabbit (inner)& $3.3^{+0.6}_{-0.5}$ & $2.3^{+0.4}_{-0.3}$ & $7.6\pm0.4$(12$\pm0.6$)
& $2.3\pm0.2$ & $1.8\pm0.14$ & \\
& $2.3^\dag$ & $1.7^{+0.1}_{-0.1}$ & $8.2\pm0.4$(12$\pm0.6$) \\

R1 & $1.0^{+2.3}_{-0.5}$ & $0.13^{+1.2}_{-0.8}$ & $1.6\pm0.3$(2.4$\pm0.5$)
& $2.5^{+0.7}_{-0.4}$ & $1.8^{+0.3}_{-0.1}$ & $1.3\pm 0.1$ ($1.8\pm0.1$)\\
& $2.5^\dag$ & $0.9^{+0.7}_{-0.8}$ & $1.4\pm 0.3$(2.1$\pm0.5$)
& \\

R2 
& $4.1^{+11}_{-2.5}$ & $1.8^\dag$ & $0.4\pm0.16$(0.6$\pm0.2$) 
& $2.0\pm0.5$ & $1.8^{+0.25}_{-0.15}$ & $0.98\pm 0.06$($1.3\pm0.1$) \\

\hline
Src 2 & $0.6^{+1.5}_{-0.6}$ & $0.4^{+0.9}_{-0.7}$ & $1.2\pm0.2$($1.7\pm0.3$)
& $1.0\pm 0.4$ & $0.8\pm0.3$ & $3.7\pm0.3$($5.2\pm0.4 $) \\
Src 3 & $3.6^{+1.9}_{-1.2}$ & $1.7^*$ & $1.2\pm0.3$($1.9\pm0.5$)
& $14.6^{+8}_{-5}$ & $3.7^{+1.6}_{-1.2}$ & $1.4\pm0.2$($6.9\pm1 $) \\
&&&
& $6.9^{+1.6}_{-1.6}$ & $1.7^*$ & $2.6\pm1.2$($5.2\pm2.4$) \\ 

\hline
Star 1 & $0.07^{+0.1}_{-0.07}$ & $0.5^{+0.07}_{-0.1}$ & $1.5\pm0.1$ \\
Star 1 (2792) & 0.07$^*$ & $0.4^{+0.05}_{-0.06}$ & $1.1\pm0.1$ \\
Star 1 (2794) & 0.07$^*$ & $0.6^{+0.04}_{-0.04}$ & $2.2\pm0.15$ \\
Star 2 & $0.5^{+0.4}_{-0.3}$ & $1.4^{+2.5}_{-0.2}$ & $0.6\pm0.06$ \\
Star 3 & $0.5^{+0.3}_{-0.2}$ & $0.6^{+0.1}_{-0.3}$ & $0.4\pm0.04$
\end{tabular}
\label{spec}
\break

$^*$fixed
\qquad
$^\dag$fixed at {\it XMM-Newton} values \\
For the \emph{Chandra} observations, rabbit nebula, Src 2 and the stars are fitted 
simultaneously using both data sets, while the other objects are fitting using on-axis 
observations only.
\qquad
$^1$in units of $10^{22}$
\qquad
$^2$for Mekal model
\qquad
$^3$absorbed flux in units of $10^{-13}$ ergs cm$^{-2}$s$^{-1}$, 0.5-8keV band, 
(unabsorbed 2-10\,keV flux); both with $1\sigma$ single parameter errors.
\end{table}

\section{Interpretation and Conclusions}

	One significant advantage of our joint {\it Chandra/XMM-Newton} analysis is that
we can isolate point sources from the diffuse X-ray nebulae while obtaining
sufficient counts for meaningful spectral fits. 
With increased sensitivity, the already multicomponent X-ray/radio
complex associated with GeV J1417$-$6100 appears even more complicated. 
In the following we adopt the more constraining of the X-ray fits 
(generally from {\it XMM-Newton}) in each case. However, Table
2 shows that even the best measured parameters have substantial uncertainty;
if errors in spectral parameters are included, the unabsorbed fluxes are 
nearly unconstrained. Accordingly we use best-fit parameters to discuss
plausible source models, but the results are illustrative rather than definitive.

	For PSR J1420$-$6048 the Cordes\&Lazio NE2001 DM model gives 
d=5.6$\pm0.8$kpc. At this distance the unabsorbed 2-10keV X-ray flux gives a
K3 PWN luminosity of $\sim 4.9 \times 10^{33} d_{5.6}^2 {\rm erg/s}$;
the pulsar is about 10\% as bright. 
Possenti {\it et al.} (2002) find that the total (PSR+PWN) flux scales as 
$L_X({\rm 2-10keV}) \approx 1.8 \times 10^{38}{\dot E}_{40}^{1.34} {\rm erg/s}$,
which predicts $L_X \approx 1.8 \times 10^{34} {\rm erg/s}$ for K3. This
would be consistent with the observed K3 flux at a distance closer to
10kpc.  For the pulsar emission itself Saito (1997) found an empirical relation 
for the 2-10keV {\it ASCA} pulsed flux: 
$L_X (2-10{\rm keV}) \approx 10^{37} {\dot E}_{40}^{1.5}{\rm erg/s}$.
Our hard (presumably magnetospheric) spectrum from PSR J1420$-$6048 corresponds to
an unabsorbed luminosity $L_X (2-10{\rm keV}) \approx 5 \times 10^{32} d_{5.6}^2{\rm erg/s}$. 
We do not detect pulsations here, but note that the light curve of RRJ01 is 
consistent with a large pulse fraction (98\% $1\sigma$ upper limit); if 60\% pulsed, our
observed PSR J1420$-$6048 point source flux would match the Saito (1997) prediction.
Recently Gotthelf (2003) has noted correlations between pulsar and PWN 
X-ray spectral indices and their luminosities. For PSR J1420$-$6048, these 
predict $\Gamma_{PSR} = 1.2$ and $\Gamma_{PWN} = 1.7$,
which are certainly consistent with our rather poorly determined values. 

	With only a handful of counts, little can be said about the
morphology of the K3 region. Indeed it is remarkable that such an
energetic pulsar does not show brighter compact PWN emission. The 
arc of X-ray emission near the pulsar (Figure 2) could be interpreted 
as a bow shock with standoff $\theta \sim 3^{\prime\prime}$.
In this case, we constrain the local density and pulsar velocity by 
$n~v_7^2 \approx 2.6/d_{5.6}^2 {\rm cm^{-3}}$ for
a velocity of 100$v_7$km/s. This is certainly compatible with
typical pulsar speeds. However, such a young $\tau_c = P/(2{\dot P})=1.3 \times 10^4$y 
pulsar is more likely still inside of its parent SNR. Indeed we might
associate this with the large scale X-ray emission in the K3 region. In
this case interpreting the arc as a termination shock implies a local
pressure $P \sim 4 \times 10^{-10} d_{5.6}^{-2} {\rm g\,cm^2/s^2}$.
This is compatible with a simple Sedov estimate of the interior
pressure $P_{int} \sim 5 \times 10^{-10} E_{51}^{2/5}n^{3/5}
\tau_4^{-6/5} {\rm g\,cm^2/s^2}$, although whether the PWN is
`crushed' depends on the detailed evolutionary state of the SNR
(van der Swaluw et al 2001).

	For the rabbit and its embedded sources we can use the
spectral fit parameters to check consistency with a PSR/PWN model.
First, given that R2 is centrally placed in the diffuse emission, that
its absorption column agrees with that of the rabbit diffuse emission
and that it shows the best (albeit still not very significant)
candidate pulse period, we tentatively assign this source as the parent pulsar
of the PWN. R1 is presumably then a background source, plausibly a faint AGN.
The {\it XMM-Newton} fit $\Gamma$ for R1 is consistent with an AGN interpretation, although
the {\it Chandra} data prefer a flatter spectral index. The modest evidence
for variability between the {\it Chandra} pointings is also consistent with
an AGN identification.

	Using Gotthelf's PWN fits, the measured spectral index of the 
rabbit $\Gamma_{PWN}=1.5 \pm 0.14$ would imply ${\rm Log}({\dot E})=36.7\pm0.2
{\rm erg/s}$ for the parent pulsar; the inner nebula $\Gamma_{PWN}=1.8$ would
imply  ${\rm Log}({\dot E})=37.2\pm0.2{\rm erg/s}$. Even for the steeper
inner rabbit spectrum, Gotthelf's correlation would imply a pulsar
spectral index of $\Gamma_{PSR}=1.4 \pm  0.2$, substantially flatter
than the value for R2.  The nominal spin parameters of \S2.2 imply
a spindown power of  ${\rm Log}({\dot E})=37.5$ with the rather young
age of $\tau_4=0.16$.  The combined 2-10keV flux of the rabbit using
the {\it XMM-Newton} spectral parameters and R2 is $\sim 7.4\times 10^{-12} {\rm erg/cm^2/s}$.
Comparing with the Possenti {\it et al.} prediction gives a distance
of $d \sim 4.4 {\dot E}_{37}^{0.76}$kpc.  Given the uncertainties, we 
scale to a fiducial d=5kpc in the following discussion.

	Certainly if the R2 pulsar candidate is as young as suggested by the nominal 
spin parameters of \S2.2, then one would expect to see evidence of its birth supernova
nearby. The faint outer emission surrounding the rabbit could represent
the birth site. In that case, the arc of emission in Figure 3 should be the 
pulsar wind termination shock; the extension to the southeast could track the 
pulsar motion or may be a channel cleared by a polar outflow, e.g. as seen for 
PSR B1509$-$58 \citet{gaet02}.

	The offset of the candidate pulsar from the center of the rabbit
and the symmetric arc surrounding it (Fig. 3), however suggest a bow shock
structure and so we comment on this possibility. To achieve a standoff
and as large as the observed $\theta \sim 25^{\prime\prime}$ at $d\sim$5kpc
requires a very low space velocity or external medium density $n v_7^2 \approx 0.05
{\dot E}_{37}/d_{5}^2$.  Thus even for a low density local ISM of
$10^{-2}n_{-2} {\rm cm^{-3}}$, we would expect the pulsar birth-site
to lie only $\sim 1.5^{\prime} ({\dot E}_{37}/n_{-2})^{1/2} \tau_4/
d_{5}^2$ away. Along the nebula axis there are several shell-like structures
in the right wing of the kookaburra at a distance of 5-10$^\prime$.
However it is difficult to reconcile supernova birth sites this far
away with a large bow shock standoff, unless the pulsar is both more energetic
and older than our fiducial values. Even then shell radii should be
$\sim 8.5 (E_{51}/n)^{1/5} \tau_4^{2/5}/d_5$arcmin, i.e. larger
than seen if the ISM density is low. Accordingly, at this stage the `termination
shock' picture seems more appealing.

	We must conclude that the complex of radio/X-ray nebulae
associated with GeV J1417$-$6100 has not yet been fully disentangled.
The case remains strong that there are two radio/X-ray PWNe in
the error contours of this source. The properties of the K3
nebula are consistent with those expected around its young
energetic pulsar, although it lacks bright compact PWN structure. 
The more puzzling `rabbit' nebula still shows all
the hallmarks of a PWN around a comparably energetic object. Indeed
our {\it Chandra/XMM-Newton} imaging has uncovered some of the PWN structure
and has identified embedded point source candidates for the pulsar
itself. Unfortunately the present observations lacked the sensitivity 
to make a definitive pulse search of these X-ray sources. Without
a secure pulse period, age and spindown luminosity, the interpretation
of the rabbit as a PWN awaits confirmation.

\acknowledgments

	This work was supported in part by NASA grants SAO G02-3083 and NAG5-13344.
We also wish to thank the referee for a very careful reading and detailed critique;
response to this tightened up the paper appreciably.


\end{document}